\def\sphinxdocclass{report}
\documentclass[a4paper,10pt,openany,oneside]{sphinxmanual}
\usepackage{iftex}

\ifPDFTeX
  \usepackage[utf8]{inputenc}
\fi
\ifdefined\DeclareUnicodeCharacter
  \DeclareUnicodeCharacter{00A0}{\nobreakspace}
\fi
\usepackage{cmap}
\usepackage[T1]{fontenc}
\usepackage{amsmath,amssymb,amstext}
\usepackage[english]{babel}
\usepackage{times}
\usepackage[Bjarne]{fncychap}
\usepackage{longtable}
\usepackage{sphinx}
\usepackage{multirow}
\usepackage{eqparbox}

\addto\captionsenglish{}
\addto\captionsenglish{}
\SetupFloatingEnvironment{literal-block}{name=Listing }

\addto\extrasenglish{}

\usepackage{amssymb}
\DeclareUnicodeCharacter{21A6}{$\mapsto$}
\DeclareUnicodeCharacter{2190}{$\leftarrow$}
\DeclareUnicodeCharacter{2194}{$\leftrightarrow$}
\DeclareUnicodeCharacter{21D2}{$\Rightarrow$}
\DeclareUnicodeCharacter{21D4}{$\Leftrightarrow$}
\DeclareUnicodeCharacter{2200}{$\forall$}
\DeclareUnicodeCharacter{2203}{$\exists$}
\DeclareUnicodeCharacter{2204}{$\nexists$}
\DeclareUnicodeCharacter{2205}{$\emptyset$}
\DeclareUnicodeCharacter{2208}{$\in$}
\DeclareUnicodeCharacter{2209}{$\notin$}
\DeclareUnicodeCharacter{2227}{$\wedge$}
\DeclareUnicodeCharacter{2228}{$\vee$}
\DeclareUnicodeCharacter{2229}{$\cap$}
\DeclareUnicodeCharacter{2260}{$\neq$}
\DeclareUnicodeCharacter{2261}{$\equiv$}
\DeclareUnicodeCharacter{2264}{$\leq$}
\DeclareUnicodeCharacter{2265}{$\geq$}
\DeclareUnicodeCharacter{2282}{$\subset$}
\DeclareUnicodeCharacter{2286}{$\subseteq$}
\DeclareUnicodeCharacter{2288}{$\nsubseteq$}
\DeclareUnicodeCharacter{228F}{$\sqsubset$}
\DeclareUnicodeCharacter{22A5}{$\perp$}

\usepackage{enumitem}
\setlist[description]{style=nextline}

\title{Multi-Party Encrypted Messaging Protocol design document}
\date{11 January 2016}
\release{0.3-2-ge104946}
\author{Mega Limited, Auckland, New Zealand \and Guy Kloss <gk@mega.co.nz>}
\newcommand{\sphinxlogo}{}
\renewcommand{\releasename}{Release}
\makeindex

\makeatletter
\def\PYG@reset{\let\PYG@it=\relax \let\PYG@bf=\relax%
    \let\PYG@ul=\relax \let\PYG@tc=\relax%
    \let\PYG@bc=\relax \let\PYG@ff=\relax}
\def\PYG@tok#1{\csname PYG@tok@#1\endcsname}
\def\PYG@toks#1+{\ifx\relax#1\empty\else%
    \PYG@tok{#1}\expandafter\PYG@toks\fi}
\def\PYG@do#1{\PYG@bc{\PYG@tc{\PYG@ul{%
    \PYG@it{\PYG@bf{\PYG@ff{#1}}}}}}}
\def\PYG#1#2{\PYG@reset\PYG@toks#1+\relax+\PYG@do{#2}}

\expandafter\def\csname PYG@tok@s1\endcsname{\def\PYG@tc##1{\textcolor[rgb]{0.25,0.44,0.63}{##1}}}
\expandafter\def\csname PYG@tok@cpf\endcsname{\let\PYG@it=\textit\def\PYG@tc##1{\textcolor[rgb]{0.25,0.50,0.56}{##1}}}
\expandafter\def\csname PYG@tok@kd\endcsname{\let\PYG@bf=\textbf\def\PYG@tc##1{\textcolor[rgb]{0.00,0.44,0.13}{##1}}}
\expandafter\def\csname PYG@tok@nl\endcsname{\let\PYG@bf=\textbf\def\PYG@tc##1{\textcolor[rgb]{0.00,0.13,0.44}{##1}}}
\expandafter\def\csname PYG@tok@vg\endcsname{\def\PYG@tc##1{\textcolor[rgb]{0.73,0.38,0.84}{##1}}}
\expandafter\def\csname PYG@tok@sb\endcsname{\def\PYG@tc##1{\textcolor[rgb]{0.25,0.44,0.63}{##1}}}
\expandafter\def\csname PYG@tok@gt\endcsname{\def\PYG@tc##1{\textcolor[rgb]{0.00,0.27,0.87}{##1}}}
\expandafter\def\csname PYG@tok@nv\endcsname{\def\PYG@tc##1{\textcolor[rgb]{0.73,0.38,0.84}{##1}}}
\expandafter\def\csname PYG@tok@si\endcsname{\let\PYG@it=\textit\def\PYG@tc##1{\textcolor[rgb]{0.44,0.63,0.82}{##1}}}
\expandafter\def\csname PYG@tok@cm\endcsname{\let\PYG@it=\textit\def\PYG@tc##1{\textcolor[rgb]{0.25,0.50,0.56}{##1}}}
\expandafter\def\csname PYG@tok@sx\endcsname{\def\PYG@tc##1{\textcolor[rgb]{0.78,0.36,0.04}{##1}}}
\expandafter\def\csname PYG@tok@sr\endcsname{\def\PYG@tc##1{\textcolor[rgb]{0.14,0.33,0.53}{##1}}}
\expandafter\def\csname PYG@tok@m\endcsname{\def\PYG@tc##1{\textcolor[rgb]{0.13,0.50,0.31}{##1}}}
\expandafter\def\csname PYG@tok@s2\endcsname{\def\PYG@tc##1{\textcolor[rgb]{0.25,0.44,0.63}{##1}}}
\expandafter\def\csname PYG@tok@gs\endcsname{\let\PYG@bf=\textbf}
\expandafter\def\csname PYG@tok@ne\endcsname{\def\PYG@tc##1{\textcolor[rgb]{0.00,0.44,0.13}{##1}}}
\expandafter\def\csname PYG@tok@o\endcsname{\def\PYG@tc##1{\textcolor[rgb]{0.40,0.40,0.40}{##1}}}
\expandafter\def\csname PYG@tok@kt\endcsname{\def\PYG@tc##1{\textcolor[rgb]{0.56,0.13,0.00}{##1}}}
\expandafter\def\csname PYG@tok@k\endcsname{\let\PYG@bf=\textbf\def\PYG@tc##1{\textcolor[rgb]{0.00,0.44,0.13}{##1}}}
\expandafter\def\csname PYG@tok@w\endcsname{\def\PYG@tc##1{\textcolor[rgb]{0.73,0.73,0.73}{##1}}}
\expandafter\def\csname PYG@tok@mb\endcsname{\def\PYG@tc##1{\textcolor[rgb]{0.13,0.50,0.31}{##1}}}
\expandafter\def\csname PYG@tok@nn\endcsname{\let\PYG@bf=\textbf\def\PYG@tc##1{\textcolor[rgb]{0.05,0.52,0.71}{##1}}}
\expandafter\def\csname PYG@tok@nc\endcsname{\let\PYG@bf=\textbf\def\PYG@tc##1{\textcolor[rgb]{0.05,0.52,0.71}{##1}}}
\expandafter\def\csname PYG@tok@mh\endcsname{\def\PYG@tc##1{\textcolor[rgb]{0.13,0.50,0.31}{##1}}}
\expandafter\def\csname PYG@tok@c1\endcsname{\let\PYG@it=\textit\def\PYG@tc##1{\textcolor[rgb]{0.25,0.50,0.56}{##1}}}
\expandafter\def\csname PYG@tok@gh\endcsname{\let\PYG@bf=\textbf\def\PYG@tc##1{\textcolor[rgb]{0.00,0.00,0.50}{##1}}}
\expandafter\def\csname PYG@tok@sh\endcsname{\def\PYG@tc##1{\textcolor[rgb]{0.25,0.44,0.63}{##1}}}
\expandafter\def\csname PYG@tok@ss\endcsname{\def\PYG@tc##1{\textcolor[rgb]{0.32,0.47,0.09}{##1}}}
\expandafter\def\csname PYG@tok@nt\endcsname{\let\PYG@bf=\textbf\def\PYG@tc##1{\textcolor[rgb]{0.02,0.16,0.45}{##1}}}
\expandafter\def\csname PYG@tok@nf\endcsname{\def\PYG@tc##1{\textcolor[rgb]{0.02,0.16,0.49}{##1}}}
\expandafter\def\csname PYG@tok@kc\endcsname{\let\PYG@bf=\textbf\def\PYG@tc##1{\textcolor[rgb]{0.00,0.44,0.13}{##1}}}
\expandafter\def\csname PYG@tok@gr\endcsname{\def\PYG@tc##1{\textcolor[rgb]{1.00,0.00,0.00}{##1}}}
\expandafter\def\csname PYG@tok@nb\endcsname{\def\PYG@tc##1{\textcolor[rgb]{0.00,0.44,0.13}{##1}}}
\expandafter\def\csname PYG@tok@err\endcsname{\def\PYG@bc##1{\setlength{\fboxsep}{0pt}\fcolorbox[rgb]{1.00,0.00,0.00}{1,1,1}{\strut ##1}}}
\expandafter\def\csname PYG@tok@go\endcsname{\def\PYG@tc##1{\textcolor[rgb]{0.20,0.20,0.20}{##1}}}
\expandafter\def\csname PYG@tok@vi\endcsname{\def\PYG@tc##1{\textcolor[rgb]{0.73,0.38,0.84}{##1}}}
\expandafter\def\csname PYG@tok@bp\endcsname{\def\PYG@tc##1{\textcolor[rgb]{0.00,0.44,0.13}{##1}}}
\expandafter\def\csname PYG@tok@kp\endcsname{\def\PYG@tc##1{\textcolor[rgb]{0.00,0.44,0.13}{##1}}}
\expandafter\def\csname PYG@tok@ow\endcsname{\let\PYG@bf=\textbf\def\PYG@tc##1{\textcolor[rgb]{0.00,0.44,0.13}{##1}}}
\expandafter\def\csname PYG@tok@gd\endcsname{\def\PYG@tc##1{\textcolor[rgb]{0.63,0.00,0.00}{##1}}}
\expandafter\def\csname PYG@tok@cs\endcsname{\def\PYG@tc##1{\textcolor[rgb]{0.25,0.50,0.56}{##1}}\def\PYG@bc##1{\setlength{\fboxsep}{0pt}\colorbox[rgb]{1.00,0.94,0.94}{\strut ##1}}}
\expandafter\def\csname PYG@tok@sc\endcsname{\def\PYG@tc##1{\textcolor[rgb]{0.25,0.44,0.63}{##1}}}
\expandafter\def\csname PYG@tok@ni\endcsname{\let\PYG@bf=\textbf\def\PYG@tc##1{\textcolor[rgb]{0.84,0.33,0.22}{##1}}}
\expandafter\def\csname PYG@tok@gp\endcsname{\let\PYG@bf=\textbf\def\PYG@tc##1{\textcolor[rgb]{0.78,0.36,0.04}{##1}}}
\expandafter\def\csname PYG@tok@na\endcsname{\def\PYG@tc##1{\textcolor[rgb]{0.25,0.44,0.63}{##1}}}
\expandafter\def\csname PYG@tok@mi\endcsname{\def\PYG@tc##1{\textcolor[rgb]{0.13,0.50,0.31}{##1}}}
\expandafter\def\csname PYG@tok@mf\endcsname{\def\PYG@tc##1{\textcolor[rgb]{0.13,0.50,0.31}{##1}}}
\expandafter\def\csname PYG@tok@vc\endcsname{\def\PYG@tc##1{\textcolor[rgb]{0.73,0.38,0.84}{##1}}}
\expandafter\def\csname PYG@tok@s\endcsname{\def\PYG@tc##1{\textcolor[rgb]{0.25,0.44,0.63}{##1}}}
\expandafter\def\csname PYG@tok@mo\endcsname{\def\PYG@tc##1{\textcolor[rgb]{0.13,0.50,0.31}{##1}}}
\expandafter\def\csname PYG@tok@no\endcsname{\def\PYG@tc##1{\textcolor[rgb]{0.38,0.68,0.84}{##1}}}
\expandafter\def\csname PYG@tok@cp\endcsname{\def\PYG@tc##1{\textcolor[rgb]{0.00,0.44,0.13}{##1}}}
\expandafter\def\csname PYG@tok@gu\endcsname{\let\PYG@bf=\textbf\def\PYG@tc##1{\textcolor[rgb]{0.50,0.00,0.50}{##1}}}
\expandafter\def\csname PYG@tok@se\endcsname{\let\PYG@bf=\textbf\def\PYG@tc##1{\textcolor[rgb]{0.25,0.44,0.63}{##1}}}
\expandafter\def\csname PYG@tok@sd\endcsname{\let\PYG@it=\textit\def\PYG@tc##1{\textcolor[rgb]{0.25,0.44,0.63}{##1}}}
\expandafter\def\csname PYG@tok@kr\endcsname{\let\PYG@bf=\textbf\def\PYG@tc##1{\textcolor[rgb]{0.00,0.44,0.13}{##1}}}
\expandafter\def\csname PYG@tok@ch\endcsname{\let\PYG@it=\textit\def\PYG@tc##1{\textcolor[rgb]{0.25,0.50,0.56}{##1}}}
\expandafter\def\csname PYG@tok@nd\endcsname{\let\PYG@bf=\textbf\def\PYG@tc##1{\textcolor[rgb]{0.33,0.33,0.33}{##1}}}
\expandafter\def\csname PYG@tok@ge\endcsname{\let\PYG@it=\textit}
\expandafter\def\csname PYG@tok@kn\endcsname{\let\PYG@bf=\textbf\def\PYG@tc##1{\textcolor[rgb]{0.00,0.44,0.13}{##1}}}
\expandafter\def\csname PYG@tok@il\endcsname{\def\PYG@tc##1{\textcolor[rgb]{0.13,0.50,0.31}{##1}}}
\expandafter\def\csname PYG@tok@c\endcsname{\let\PYG@it=\textit\def\PYG@tc##1{\textcolor[rgb]{0.25,0.50,0.56}{##1}}}
\expandafter\def\csname PYG@tok@gi\endcsname{\def\PYG@tc##1{\textcolor[rgb]{0.00,0.63,0.00}{##1}}}

\makeatother

\begin{document}

\maketitle
\tableofcontents
\phantomsection\label{index::doc}

\newpage
\begin{DUlineblock}{0em}
\item[] © 2015 Mega Limited, Auckland, New Zealand.
\item[] \url{https://mega.nz/}
\end{DUlineblock}\vspace{-\baselineskip}

This work is licensed under the Creative Commons
Attribution-ShareAlike 4.0 International License. For the full text of the license in various formats, and other
details, see \url{https://creativecommons.org/licenses/by-sa/4.0/}

To give a non-legal human-readable summary of (and \emph{not} a substitute for) the
license, you are free to:
\begin{itemize}
\item {} 
Share -- copy and redistribute the material in any medium or format

\item {} 
Adapt -- remix, transform, and build upon the material

\end{itemize}

for any purpose, even commercially. The licensor cannot revoke these freedoms
as long as you follow the license terms:
\begin{itemize}
\item {} 
Attribution -- You must give appropriate credit, provide a link to the
license, and indicate if changes were made; but not in any way that suggests
the licensor endorses you or your use.

\item {} 
ShareAlike -- If you distribute your adaptions of the material, you must do
this under the same license as the original. You may not apply additional
restrictions when doing so, legal or technical.

\end{itemize}

No warranties are given. Other legal rights may extend (e.g. fair use and fair
dealing) or restrict (e.g. publicity, privacy, or moral rights) your permission
to use this material, outside of the freedoms given by this license.

\chapter{Strongvelope Multi-Party Encrypted Messaging Protocol}
\label{preface:strongvelope-multi-party-encrypted-messaging-protocol}\label{preface::doc}\label{preface:strongvelope-documentation}
In this document we describe the design of a multi-party messaging
encryption protocol ``Strongvelope''.  We hope that it will prove useful
to people interested in understanding the inner workings of this
protocol as well as cryptography and security experts to review the
underlying concepts and assumptions.

In this design paper we are outlining the perspective of chat message
protection through the Strongvelope module. This is different from the
\emph{product} (the Mega chat) and the transport means which it will be
used with. Aspects of the chat product and transport are only referred
to where appropriate, but are not subject to discussion in this
document.

\section{Intent}
\label{preface:intent}
The Strongvelope protocol is intended to protect the privacy and
confidentiality of content exchanged in Mega's chat application among
groups of participants.

As underlying design goals a set of security properties needs to be
met (see next section). These have to be upheld in instant as well as
offline messaging.  Furthermore, upon changes of the set of chat
participants, it needs to be ensured that the used encryption keys
change.  It is required that newly added participants are unable to
decrypt previously sent messages, and former (now excluded)
participants are unable to decrypt messages after their exclusion.

\section{Security Properties}
\label{preface:security-properties}
Our design rationale was based on the need of implementing a
multi-party capable instant messaging protocol, that would be light
enough to be executed over the Internet/World Wide Web when
implemented in a JavaScript text chat client or on mobile devices.
The protocol would need to provide primarily confidentiality of the
conversation as well as message and message origin authenticity.
\begin{itemize}
\item {} \begin{description}
\item[{Confidentiality}] \leavevmode
of the conversation, so its content is not accessible or readable
by an outsider.

\end{description}

\item {} \begin{description}
\item[{Message authenticity}] \leavevmode
against alteration during transport or storage.  A recipient can be
assured that the message is arriving exactly as it has been
originally authored.

\end{description}

\item {} \begin{description}
\item[{Message origin authenticity}] \leavevmode
against both outsider intrusion and the impersonation of existing
participants by other malicious participants in the session.  This
means that the user can be assured of the authenticity of the
sender of each original message even if other participants in the
room try to impersonate the sender and send messages on their
behalf.

\end{description}

\end{itemize}

\section{Scope and Limitations}
\label{preface:scope-and-limitations}
Any further assurances underpinned by cryptography are not (yet)
supported by this protocol.  Further security enhancing properties may
be introduced in the future evolution of Strongvelope or are part of a
different encryption module featuring a different (but overlapping)
set of properties (such as the mpENC protocol \phantomsection\label{preface:id1}{\hyperref[references:mpenc]{\crossref{{[}mpENC{]}}}}).

\section{Assumptions}
\label{preface:assumptions}
The design of Strongvelope was driven by a number of constraining
assumptions.

\subsection{Underlying Transport}
\label{preface:underlying-transport}
Strongvelope is intended to be a protocol design that is agnostic of
the used transport protocol.  However, we are presuming that the
underlying transport mechanism message will retain message order.

\subsection{Chat ``Rooms''/Transport Channels}
\label{preface:chat-rooms-transport-channels}
Group chats require a shared transport channel, often referred to as a
``room'' in which the participants of the chat gather and receive
messages from the group.  We are assuming that there is at the most
\emph{one} Strongvelope multi-party chat present per transport channel
(room).  Not all members of the channel need to be participants in the
Strongvelope chat session.  This is intended, as members may (more or
less freely) be added to or removed from participation in a channel,
but they explicitly need to be added to, or removed from the
Strongvelope participants in a session.  Therefore it is expected that
there are potentially a few extra members in the chat room, which are
(not or not yet) participating in the encrypted session.

Note that in this context a ``chat room'' or a ``channel'' are synonymous
for the underlying technical message exchange mechanism used for the
group communication. In contrast a ``chat'' is the actual exchange of
messages on top of the room or channel.

\subsection{Transport Meta-Data}
\label{preface:transport-meta-data}
It is assumed that the transport mechanism will carry a limited amount
of meta-data for message delivery, which is to be used for the
encryption protocol to work. This includes specifically a sender
(originator) ID for each message as well as a unique ordering criteria
(enforced by the server) to place all messages of a single chat room
within a total order as seen by the server.

\subsection{Implementation}
\label{preface:implementation}
To make the implementation robust for working on many types of
transports, some precaution should be taken when processing incoming
protocol messages.  For example, due to the broadcast nature of some
protocols (e.g. multi-party chat rooms or IRC channels), there may not
be the possibility for directed message delivery, therefore an
implementation may adopt the option to abort processing of messages
not intended for the client itself.  Specifically, these may be
messages not intended for oneself sent over a broadcast channel
(e.g. a message sent by oneself or a message received after quitting
participation while still a member of the chat ``room'' on the transport
implementation).

\section{Messages}
\label{preface:messages}
It is not intended for Strongvelope to co-exist with clear-text
messaging.  It is assumed that \emph{all} messages will be protected
consistently via the Strongvelope protocol.  All messages contain
content required to drive the cryptographic protection, and they may
also contain an encrypted data payload.  All content in Strongvelope
messages (besides the first byte, indicating the protocol version
used) is cryptographically signed.

\section{Terminology}
\label{preface:terminology}
In this document the term ``message'' is used frequently.  To avoid
ambiguity, a ``message'' is considered to be a unit of information
transported by means of any suitable wire protocol.  Such a message
includes additional information and meta-data, such as cryptographic
keys, signatures, etc.  A user content message in the general sense is
authored by a user/participant of the chat, and is termed to be a
``message payload'' or just ``payload''.  Therefore it is possible to send
a message without a payload, which is also termed to be a ``blind
message'', as it will not need to be displayed within a client.

\chapter{Cryptographic Primitives}
\label{crypto_primitives:cryptographic-primitives}\label{crypto_primitives:crypto-primitives}\label{crypto_primitives::doc}
Mega provides a cloud-based platform enjoying a large popularity.
Therefore, server-side scalability is of importance as well as
feasibility for the messaging concept and the client implementation.
Even though server scalability tends to be orthogonal to messaging
encryption protocols, they have shown to add a significant overhead on
the server infrastructure (due to the number of ``blind'' protocol
bootstrapping messages and increased message size), so that fewer
users can be served per server provided.

Out of experience, most of the clients will be using the messenger on
the Mega platform through a Web or mobile client with limited
computational capabilities (e.g. with end-to-end cryptography
implemented in JavaScript).  To maximise user experience through a
fast response time (less lag) and reduced load on the executing end
point hardware (which often are mobile devices), it is desirable to
avoid frequent ``heavy'' computing operations (e.g. frequent
exponentiation of big integers).

The cryptographic primitives have generally been chosen to match a
general security level of 128 bit of entropy on symmetric ciphers such
as AES.  This is equivalent to 256 bit key strength on elliptic curve
public-key ciphers and 3072 bit key strength on discrete logarithm
problem based public-key ciphers (e.g. RSA, DSA).  However, 3072 bit
key strength is considered to be too expensive in many cases
(computationally as well as with its demand on the entropy source).
For security reasons NIST standardised ECC curves have been avoided\phantomsection\label{crypto_primitives:id1}{\hyperref[references:nist\string-ecc\string-failure]{\crossref{{[}NIST-ECC-Failure{]}}}}.

\section{Key Pair for Authentication}
\label{crypto_primitives:key-pair-for-authentication}
Authentication of the sender as well as the message transmitted is
performed via a cryptographic signature.  For computational efficiency
and good implementation across the different environments a signature
key pair using EdDSA with the Edwards curve 25519 (``Ed25519'')
\phantomsection\label{crypto_primitives:id2}{\hyperref[references:ed25519]{\crossref{{[}Ed25519{]}}}} is used.  It is comparably compact (256 bit public keys,
512 bit signature size, 256 bit private key seed size).

\section{Key Agreement}
\label{crypto_primitives:key-agreement}
To agree on a symmetric encryption key between a sender and recipient
a Diffie-Hellman (DH) scheme is used.  To be more easily suitable for
offline messaging capability, static ``chat keys'' using the Montgomery
curve 25519 (``Curve25519'') \phantomsection\label{crypto_primitives:id3}{\hyperref[references:curve25519]{\crossref{{[}Curve25519{]}}}} are used.  Pair-wise
encryption keys (between sender and receiver) are derived using ECDH
(elliptic curve Diffie-Hellman).

\section{(Sender) Key Encryption}
\label{crypto_primitives:sender-key-encryption}
Each sender's key is encrypted with 128 bit AES in
cipher-block-chaining (CBC) mode.  No padding is used (all encryption
keys a sender is using are exactly the size of a block).

\section{Message Authentication}
\label{crypto_primitives:message-authentication}
Each message is signed cryptographically for sender as well as message
authenticity.  To avoid message inflation and computational strain on
the clients, an elliptic curve (EC) signature scheme is used.  For
efficiency, security and reputation the Edwards curve 25519
(``Ed25519'') \phantomsection\label{crypto_primitives:id4}{\hyperref[references:ed25519]{\crossref{{[}Ed25519{]}}}} has been chosen (256 bit public keys, 512 bit
signature size, 256 bit private key seed size) providing EdDSA
signatures.

\section{Message Encryption}
\label{crypto_primitives:message-encryption}
Messages are encrypted with 128 bit AES in counter (CTR) mode.

\chapter{Message Encryption}
\label{message_encryption:message-encryption}\label{message_encryption::doc}\label{message_encryption:id1}
Every sender is responsible for generating their own symmetric
encryption key, ensuring user-controlled keys for any encrypted
content sent.  Therefore, each sender generates their own symmetric
encryption keys used for encrypting the message payload.  These
``sender keys'' then need to be exchanged with all other participants
within a chat (pair-wise).  The sender keys as well the message
payload content then need to be encrypted for message transport and
storage.

\section{Sender Key Exchange}
\label{message_encryption:sender-key-exchange}
For this purpose ``keyed'' messages exist, which will carry the sender
key to all recipients.  Each chat participant is in possession of a
Curve25519 key pair, with the public portion of this pair available to
all other participants via an API (not part of this document).  The
new sender key is encrypted to \emph{each} other participant by a key
derived through ECDH \footnote[1]{\sphinxAtStartFootnote%
Note on older legacy clients: Some users may not have used a
client recently that will generate a Curve25519 key pair. To
not undermine the user experience, sender keys to such users
will be encrypted with their RSA public keys. Upon logging
into a client with a working Strongvelope implementation a
Curve25519 key pair for the chat will be generated and used
from then on.
}.  Each recipient is able to extract and
decrypt this embedded sender key of a participant.  Each client is
responsible of tracking these user-specific sender keys, identified
uniquely by a tuple \((participant, key ID)\).

A message payload then is encrypted with this particular sender key
and embedded within the message.  Future follow-up messages only need
to state the ID of the key previously transferred to indicate which
encryption key was used to protect the message.

\subsection{Sender Keys}
\label{message_encryption:sender-keys}
Each sender key is 128 bit long for use with AES in CTR mode.

\subsection{Key IDs}
\label{message_encryption:key-ids}
Key IDs are 32 bit long, need to be unique for each sender, and
strictly monotonously increasing.  For the purpose of the
implementation they contain a 16 bit (high) portion derived from a
UNIX Epoch time stamp to the granularity of days (incremented each UTC
day), combined with a 16 bit (low) portion of a counter (starting from
zero).  The client implementation must prevent collisions or
roll-overs.

\subsection{Key Rotation}
\label{message_encryption:key-rotation}
Whenever a new sender key is required, the sender will generate one,
and send it (encrypted to all participants) along with the new key ID
to all participating recipients in a keyed message.  This is desirable
to refresh a sender key (preventing extensively long use \footnote[2]{\sphinxAtStartFootnote%
Our implementation rotates a sender key every 16 sent
messages.
}), or
when the composition of the group chat has changed (added and/or
removed participants).  Upon changes in the group composition the
first message a client sends to the group chat \emph{must} be a keyed
message stating a new sender key.

For convenience (e. g. when loading the chat history in reverse
order), the previously used key with its key ID are re-sent to
previous participants in the group as well.  The client \emph{must not}
send the previous key to newly joined participants, and \emph{must not}
send a new key to departed participants.

\subsection{Key Re-Sending}
\label{message_encryption:key-re-sending}
Sometimes the client needs to access messages out of context of a
current message exchange flow, for example when loading previous
messages from the chat history \footnote[3]{\sphinxAtStartFootnote%
Our implementation loads chat history messages in batches of
32 messages.
}.  In those cases, a message's
payload may only be decrypted with knowledge of the sender key used.
Each message does carry the sender key ID used, but the (encrypted)
key itself may be contained in one of the previous messages only.  To
ease the access to previous sender keys used, clients will re-send
their current sender keys (including the respective key ID) in regular
intervals.  Ideally, this interval \footnote[4]{\sphinxAtStartFootnote%
Our implementation re-sends sender keys every 30 total
messages (received and sent).
} is balanced with the
intervals used for key rotation \footnotemark[2] and batch size for chat history
loading \footnotemark[3].  A client may face the situation that key re-sending
is due, but no chat message is to be sent.  Therefore such a message
can be sent without a user contributed payload resulting in a ``blind
message'' (not to be displayed by the client application).

\section{Content Encryption}
\label{message_encryption:content-encryption}
Individual message content components need to be protected through
encryption.  When using a symmetric key several times, a different
initialisation vectors (IVs) or nonces \emph{must} be used.  Each message
carries one such ``message nonce'' encoded (but not encrypted) within
the message.

\subsection{Sender Key Encryption}
\label{message_encryption:sender-key-encryption}
Sender keys are encrypted using AES in CBC mode, using an
initialisation vector (IV) along with the ECDH shared secret \footnotemark[1]
with each participant.

To derive the shared secret one's own private (\(S_{own}\)) and
the other participant's public key (\(S_{own}\)) is used.  Through
Curve25519 Diffie-Hellman scalar multiplication (ECDH) and subsequent
application of a key derivation function (KDF, specifically
\phantomsection\label{message_encryption:id9}{\hyperref[references:hkdf]{\crossref{{[}HKDF{]}}}}-SHA256) the key is derived.  It is trimmed to the required key
size (128 most significant bits).  According to RFC-5869 (\phantomsection\label{message_encryption:id10}{\hyperref[references:hkdf]{\crossref{{[}HKDF{]}}}}) as
a context info string the byte sequence ``\code{strongvelope pairwise
key}'' is used, followed by the byte \code{0x01}.
\begin{equation*}
\begin{split}K_{DH, dest} = KDF(ECDH(S_{own}, P_{other}))\end{split}
\end{equation*}
To derive an IV (initialisation vector) for a recipient, the Mega user
handle of the recipient is base64 URL decoded, yielding an 8 byte (64
bit) sequence (\(u\)).  From these bytes then a keyed-hash message
authentication code (HMAC, specifically HMAC-SHA-256) is computed
using the message's master nonce (\(n\)) as a key, and
subsequently trimmed to the required IV size (128 most significant
bits).
\begin{equation*}
\begin{split}IV_{dest} = HMAC(n_{master}, u_{other})\end{split}
\end{equation*}

\subsection{Payload Encryption}
\label{message_encryption:payload-encryption}
Message payloads are encrypted using AES in CTR mode, using the sender
key and message nonce derived via computing an HMAC using the
message's master nonce as a key and the byte sequence ``\code{payload}'' as
a value.  The message nonce will be trimmed to use the 96 most
significant bits (12 bytes) only, leaving 32 bits for the counter.
\begin{equation*}
\begin{split}n_{message} = HMAC(n_{master}, \mathtt{"payload"})\end{split}
\end{equation*}

\chapter{Protocol Encoding}
\label{protocol_encoding:protocol-encoding}\label{protocol_encoding::doc}\label{protocol_encoding:id1}
All fields in the messages exchanged are encoded as TLV (type, length,
value) records.  The entire message is prepended by a single byte
indicating the protocol version (in case of future changes).  At the
moment the protocol version is \code{0x00}.

\section{TLV Types}
\label{protocol_encoding:tlv-types}
TLV records do not need to be in order according to the TLV type
numbers, as long as it is assured that the \code{SIGNATURE} record is
preceding all others..  Individual records may be missing or repeated
multiple times.  The currently defined TLV record types are:
\begin{description}
\item[{Type \code{0x01} (1): \code{SIGNATURE}}] \leavevmode
Signature for all following bytes.

\item[{Type \code{0x02} (2): \code{MESSAGE\_TYPE}}] \leavevmode
Type of message sent.

\item[{Type \code{0x03} (3): \code{NONCE}}] \leavevmode
``Message nonce'' used for encryption (individual nonces are derived
from it).

\item[{Type \code{0x04} (4): \code{RECIPIENT}}] \leavevmode
Recipient of message.  This record is repeated for all recipients
of the message.

\item[{Type \code{0x05} (5): \code{KEYS}}] \leavevmode
Message encryption keys, encrypted to a particular recipient.  This
may contain two (concatenated) keys.  The second one (if present)
is the previous sender key.  Records of this type need to occur
with the same number of records in the same order as \code{RECIPIENT}.

\item[{Type \code{0x06} (6): \code{KEY\_IDS}}] \leavevmode
Sender encryption key IDs used (or set) in this message.  If some
\code{KEYS} records will contain a second (previous) key ID, the
previous key ID needs to be given (concatenated) as well.

\item[{Type \code{0x07} (7): \code{PAYLOAD}}] \leavevmode
Encrypted payload of message.

\item[{Type \code{0x08} (8): \code{INC\_PARTICIPANT}}] \leavevmode
Participant to be included with this message.

\item[{Type \code{0x09} (9): \code{EXC\_PARTICIPANT}}] \leavevmode
Participant to be excluded with this message.

\item[{Type \code{0x0a} (10): \code{OWN\_KEY}}] \leavevmode
Own message encryption (sender) key. This is usually not required,
but will be used if legacy RSA encryption of sender keys is used for
at least one recipient. This is required to access one's own sender
key later when re-reading own chat messages later from history.

\end{description}

\section{Message Signatures}
\label{protocol_encoding:message-signatures}
All message signatures are detached cryptographic signature, signing
the entire following (encoded) binary message content. To authenticate
the entire message content, the signature \emph{must} be contained in the
first TLV record. All message signatures are computed using the
sender's Ed25519 identity key.

The content to sign is computed as follows:
\begin{equation*}
\begin{split}(magic\:number || content\:to\:sign)\end{split}
\end{equation*}
Here, \(magic\:number\) is a fixed string to distinguish the
authenticator from any other content (for now it is the byte sequence
``\code{strongvelopesig}'').

\section{Message Types}
\label{protocol_encoding:message-types}\label{protocol_encoding:id2}
The message type is indicated by a single byte.  Currently these types
are in use:
\begin{description}
\item[{Type \code{0x00} (0): \code{GROUP\_KEYED}}] \leavevmode
Message containing a sender key (initial, key rotation, key re-send).

\item[{Type \code{0x01} (1): \code{GROUP\_FOLLOWUP}}] \leavevmode
Message using an existing sender key for encryption.

\item[{Type \code{0x02} (2): \code{ALTER\_PARTICIPANTS}}] \leavevmode
Alters the list of participants for the group chat (inclusion and
exclusion).

\end{description}

\section{Keyed Messages}
\label{protocol_encoding:keyed-messages}
After the mandatory byte for the protocol version, all keyed messages
must contain the following records for a group chat of \(n\)
participants:
\begin{itemize}
\item {} 
(0x01) \code{SIGNATURE}

\item {} 
(0x02) \code{MESSAGE\_TYPE}

\item {} 
(0x03) \code{NONCE}

\item {} 
(0x04) \code{RECIPIENT} {[}\((n-1)\) records of this type{]}

\item {} 
(0x05) \code{KEYS} {[}\((n-1)\) records of this type, corresponding order
to \code{RECIPIENT} records{]}

\item {} 
(0x06) \code{KEY\_IDS}

\item {} 
(0x07) \code{PAYLOAD} {[}may only be omitted on ``blind'' messages, when
re-sending keys{]}

\end{itemize}

\section{Followup Messages}
\label{protocol_encoding:followup-messages}
Followup messages are designed to be leaner (in terms of
storage/transport size), and therefore are missing some of the TLV
records.  Messages contain the following TLVs (in the given order):
\begin{itemize}
\item {} 
(0x01) \code{SIGNATURE}

\item {} 
(0x02) \code{MESSAGE\_TYPE}

\item {} 
(0x03) \code{NONCE}

\item {} 
(0x06) \code{KEY\_IDS} {[}contains only current/used key ID{]}

\item {} 
(0x07) \code{PAYLOAD}

\end{itemize}

\section{Alter Participant Messages}
\label{protocol_encoding:alter-participant-messages}
When altering the group composition of the chat (inclusion and/or
exclusion of participants), the sender key must be rotated and the
message must distribute the new sender key to all new participants.
After the mandatory byte for the protocol version, all participant
change messages must contain the following records for a group chat of
\(n\) participants:
\begin{itemize}
\item {} 
(0x01) \code{SIGNATURE}

\item {} 
(0x02) \code{MESSAGE\_TYPE}

\item {} 
(0x03) \code{NONCE}

\item {} 
(0x04) \code{RECIPIENT} {[}\((n-1)\) records of this type{]}

\item {} 
(0x05) \code{KEYS} {[}\((n-1)\) records of this type, corresponding order
to \code{RECIPIENT} records{]}

\item {} 
(0x06) \code{KEY\_IDS}

\item {} 
(0x08) \code{INC\_PARTICIPANT} {[}a TLV record for each new participant to
include{]}

\item {} 
(0x09) \code{EXC\_PARTICIPANT} {[}a TLV record for each former participant to
exclude{]}

\item {} 
(0x07) \code{PAYLOAD} {[}may only be omitted on ``blind'' messages, when
re-sending keys{]}

\end{itemize}

\section{Legacy Sender Key Encryption}
\label{protocol_encoding:legacy-sender-key-encryption}
In case at least one recipient of a keyed message (\code{GROUP\_KEYED} or
\code{ALTER\_PARTICIPANTS} message types) requires encryption of a sender
key using the recipient participant's RSA public key, the sender must
add an additional TLV record of the \code{OWN\_KEY} type. This \code{OWN\_KEY}
record is identical to a \code{KEYS} record, with the only difference
that the key(s) are to be encrypted using one's own Curve25519 public
key to derive a shared secret. This is required as one does not have
access to a recipient's RSA private key to recover a sender key
encrypted with it (e.g. when loading transport encrypted messages
from the chat server's history).

\section{Sender Keys to Include}
\label{protocol_encoding:sender-keys-to-include}
When loading the history messages from a chat server on demand
(``backwards'' in history) one cannot easily decrypt preceding messages
before a key rotation, until the next prior key rotation message has
been retrieved as well. To ease this process, a client encrypts the
new sender key to the recipient as well as the previous sender key (if
possible and the recipient is entitled to it).

When including previous sender keys, also the \code{KEY\_IDS} record must
also include the previous sender key's ID (concatenated to the current
sender key ID). Additionally the sending client \emph{must} ensure that
only entitled participants will receive this previous sender key. For
example newly included participants are not entitled access to the
content of previous messages, and therefore \emph{must not} receive such
previous sender keys.

\chapter{Requirements for Messaging Workflows}
\label{messaging_workflow:messaging-workflow}\label{messaging_workflow:requirements-for-messaging-workflows}\label{messaging_workflow::doc}
To assure proper operation of messaging workflows among participants,
especially when encountering inclusions and/or exclusions of
participants in the chat room, some common procedures must be ensured
by the client.

Regarding sender keys to be included in keyed messages
(\code{GROUP\_KEYED} or \code{ALTER\_PARTICIPANTS} message types) one \emph{must}
follow the requirements as outlined for alter participant messages
(see {\hyperref[messaging_workflow:alter\string-participant\string-messages]{\crossref{\DUrole{std,std-ref}{Alter Participant Messages}}}}).

\section{Initialising Chat Room Participation}
\label{messaging_workflow:initialising-chat-room-participation}
When a client starts to participate in a chat session, two scenarios
are possible: The chat session could be new for the client who has
never participated in it, or the client could resume a chat it has
previously participated in.
\begin{itemize}
\item {} 
In the former case (a new chat), the client must generate a new
sender key and a corresponding key ID before sending the first
message (see {\hyperref[messaging_workflow:update\string-sender\string-key]{\crossref{\DUrole{std,std-ref}{Update Sender Key}}}}). The first message then must
then contain this sender key encrypted to all participants (keyed
message).

\item {} 
In the latter case (resuming a chat), the client must be ``seeded''
with chat history messages (see {\hyperref[messaging_workflow:seed\string-encryption\string-handler]{\crossref{\DUrole{std,std-ref}{Seed Encryption Handler}}}}) to
identify and extract own previous sender keys, as well as
participants' sender keys along with their key IDs. In case the own
client has never sent a message in the chat before, one reverts to
the ``new chat'' behaviour outlined above (see also
{\hyperref[messaging_workflow:update\string-sender\string-key]{\crossref{\DUrole{std,std-ref}{Update Sender Key}}}}).

\end{itemize}

\section{Update Sender Key}
\label{messaging_workflow:update-sender-key}\label{messaging_workflow:id1}
When updating a sender key, a user's encryption handler generates a
new key ID and associated sender key. With the next message sent, this
new sender key and key ID are sent to the participants of the chat (a
keyed message). If a previous sender key is available as well, it will
also be included to those participants only who are entitled to it
(participants in the chat when that sender key was in use). See
{\hyperref[messaging_workflow:alter\string-participant\string-messages]{\crossref{\DUrole{std,std-ref}{Alter Participant Messages}}}} for the case of a changed chat group
composition.

\section{Seed Encryption Handler}
\label{messaging_workflow:seed-encryption-handler}\label{messaging_workflow:id2}
When resuming a chat (starting to use a non-new chat, e.g. by
responding to a previously sent initial chat message), the encryption
handler needs to be ``seeded'' with sender keys. These include one's own
sender keys and key IDs. For seeding, the handler will be given a
suitably sized batch of messages from the message history to extract
sender keys and their corresponding key IDs. The handler then will
need to provide feedback on whether one's own (latest) sender key was
contained in it. In the case it was \emph{not}, a further (older,
adjoining) batch of messages need to be passed to the seeder. In the
case no own sender key could be found in the entire history available,
the client must generate a new sender key for further use (see
{\hyperref[messaging_workflow:update\string-sender\string-key]{\crossref{\DUrole{std,std-ref}{Update Sender Key}}}}).

A client may consider to ``rotate'' one's own sender key immediately
at the beginning of a session (see {\hyperref[messaging_workflow:rotate\string-key]{\crossref{\DUrole{std,std-ref}{Rotate Key}}}}).

\section{Rotate Key}
\label{messaging_workflow:id3}\label{messaging_workflow:rotate-key}
After a while of usage, a client may opt to ``rotate'' its sender
key. That is, it generates a new one with a new key ID, and sends out
a keyed message as the next message sent to the chat, informing all
participants of the new sender key in use. This strategy is also used
for the case of altering participants in a group chat. It ensures that
previous members will be disallowed any access to newly sent messages,
as well as new members are disallowed access to previous messages on a
cryptographic level (see {\hyperref[messaging_workflow:alter\string-participant\string-messages]{\crossref{\DUrole{std,std-ref}{Alter Participant Messages}}}}).

\section{Key Reminders}
\label{messaging_workflow:key-reminders}\label{messaging_workflow:id4}
In situations with long one-sided communication, the history may be
flooded with messages from one participant, and key rotation messages
of oneself or other members may be scarce. This makes the seeding
process (see {\hyperref[messaging_workflow:seed\string-encryption\string-handler]{\crossref{\DUrole{std,std-ref}{Seed Encryption Handler}}}}) difficult, as large
amounts of chat history may need to be fetched to successfully seed an
encryption handler.  Therefore, clients can send a keyed key reminder
message in (regular) intervals. These messages may also be ``blind''
messages without content, just injecting the sender key material into
the chat history again to aid future resumption.

\section{Alter Participant Messages}
\label{messaging_workflow:id5}\label{messaging_workflow:alter-participant-messages}
The process triggered by alter participant messages is a bit more
involved. In an alter participant message, one or more participants
can included into or excluded from the group chat room. A client
sending such a message will alter its local current group composition
set, and include a list of participants to be included and excluded
from the chat. Before sending such a message, the client rotates its
sender key, and notifies the newly legitimate participants of the chat
of this new sender key. The sender needs to ensure that the previous
sender key \emph{will not} be encrypted to newly included participants.

However, the recipients of an alter participant message also need to
take action upon receiving it. They need to maintain a set for
participants to include as well as to exclude on top of the current
list of current participants. When they are to send their next own
message, they need to rotate their sender key as well, and encrypt it
to the now legitimate group chat participants similarly to the sender
of the alter participant message. In the same fashion, included
participants must not be informed of the previous sender key, and
excluded participants must not receive any sender keys at all with
that message.

Due to the fact that a number of alter participant messages may be
received before the next own message is sent, the recipient needs to
update their sets for participants to include and to exclude until
such a first message is sent again. Once an own keyed message has been
sent, the client may clear their include and exclude participant sets
as well as their set of current group chat participants.

\chapter{Strongvelope Evolution}
\label{future_work:strongvelope-evolution}\label{future_work::doc}
Strongvelope is not finished in its development/evolution, yet.  The
following are some notes on aspects still to be addressed.
\begin{itemize}
\item {} 
Concept of a ``room operator'' who has moderator privileges, and thus
may be entitled for privileged operations (such as inclusion or
exclusion of participants, setting of room topics/titles, etc.)

\item {} 
...

\end{itemize}

\renewcommand{\indexname}{Index}
\printindex
\end{document}